# The tip-sample water bridge and light emission from scanning tunnelling microscopy

Michael G Boyle, J Mitra and Paul Dawson
Centre for Nanostructured Media, School of Mathematics and Physics, Queen's University, Belfast BT7 1NN, UK
Email: j.mitra@qub.ac.uk

Light emission spectrum from a scanning tunnelling microscope (LESTM) is investigated as a function of relative humidity and shown to be a novel and sensitive means for probing the growth and properties of a water meniscus in the nm-scale. An empirical model of the light emission process is formulated and applied successfully to replicate the decay in light intensity and spectral changes observed with increasing relative humidity. The modelling indicates a progressive water filling of the tip-sample junction with increasing humidity or, more pertinently, of the volume of the localized surface plasmons responsible for light emission; it also accounts for the effect of asymmetry in structuring of the water molecules with respect to polarity of the applied bias. This is juxtaposed with the case of a non-polar liquid in the tip-sample nano cavity where no polarity dependence of the light emission is observed. In contrast to the discrete detection of the presence/absence of water bridge in other scanning probe experiments by measurement of the feedback parameter for instrument control LESTM offers a means of continuously monitoring the development of the water bridge with sub-nm sensitivity. The results are relevant to applications such as dip-pen nanolithography and electrochemical scanning probe microscopy.

## 1. Introduction

There have been significant advances and novel adaptations to the probe that started off as the scanning tunnelling microscope (STM) and has evolved to define a general class of techniques that we today collectively refer to as scanning probe microscopy (SPM). While a great deal of fundamental scientific work has issued from the use of SPM under ultrahigh vacuum conditions the majority day-to-day use of the technique occurs under 'ambient' conditions with limited environmental control. One of the key issues in ambient SPM is the role of water in the tip-sample gap. This is of importance in biological [1] and many electrochemical [2, 3] applications where water forms the ambient medium. The capillary condensation of a nano-scale water droplet in the gap between tip and sample, the focus of the present investigation, is of fundamental importance in ambient instrument operation and imaging in general. This is a matter of both long-running [4] and topical interest [5, 6] in the context of atomic force microscopy, especially in applications such as dip pen lithography [7]. A tip–sample water bridge also plays a critical role in instrument operation in configurations in which tip-sample distance regulation relies on the detection of a shear force between a laterally dithered tip and the sample a scheme that is widely adopted in scanning near-field optical microscopy [8,9]. The detailed mechanism of water bridge formation in SPM has also been addressed though molecular dynamics simulations [10].

In the present study we utilize the phenomenon of light emission from scanning tunnelling microscope (LESTM) [11] and its associated spectroscopy [12-16] to study condensation of water in the tip-sample gap. In essence this is the converse situation to tip-enhanced Raman scattering [17] and apertureless near-field microscopy [9] where light is fed into the tip-sample gap and the findings here are pertinent to those applications. In examining LESTM, the formation and subsequent growth of a water meniscus is achieved through using relative humidity (RH) as the external control.

Figure 1 offers a schematic of the mechanism of LESTM for metal-metal junctions, extending to the effect of a water bridge in the tip-sample gap. For the metal-metal set-up the origin of the light emission is the tunnelling electron excitation and subsequent radiative decay of localized surface plasmon (LSP) modes supported in the tip-sample nano-cavity [11, 12]. The emission is primarily governed by (i) the optical frequency fluctuations in the tunnel current (which in turn depends on tip and sample density of states) (ii) the dielectric properties of the tip, sample and the surrounding medium and (iii) the geometry of the nano-cavity, notably the tip profile. LSP excitation at discrete energies is manifest in spectral peaks in the light emission. The detailed theoretical understanding of LESTM has been largely based on corresponding analyses [18, 19] of the phenomenon of light emission from tunnelling in metal-oxide-metal junctions [20, 21]. The crucial departure has been the replacement of modelling the grain size/distribution [18] of the counter electrode in a metal-oxide-metal device with that of the STM tip profile and introduction of the STM tunnel current formalism [12, 22-24].

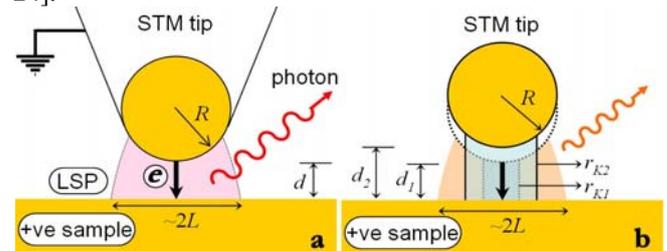

Figure 1 (a) Schematic illustration of the light emission process from a STM with a tip of end radius $R$ and a positively biased sample, separated by tunnel gap $d$. The LSP volume is depicted by $L \sim \sqrt{(Rd)}$. (b) With increasing RH more water condenses in the tunnel gap increasing the water column radius from $r_{K1}$ to $r_{K2}$, thus increasing the total tip–sample current (see text). This causes the (constant-current) feedback of the STM to withdraw the tip from separation $d_1$ to $d_2$, resulting is a decrease in the emission intensity and a blue shift in the spectral peak.

A key point in this investigation is that while the emitted light serves as a highly sensitive probe of the tip-sample dielectric environment it is not the physical parameter that is used in instrument feedback for tip control. Moreover, the originating excitation, the LSP, in general occupies a greater (nano-scale) volume than the bridging water droplet, certainly in the initial stages of the droplet's growth, as depicted in Figure 1. This means that the connecting water bridge does not predominantly or exclusively channel the electromagnetic interaction between tip and sample, thus its properties can be monitored in quasi-continuous fashion until its volume exceeds that of the LSP modal volume. (Analysis of the emission spectra offers a route to the optical characterization of liquid bridges down to the zepto- ($10^{-21}$) liter scale). There is thus significant contrast with other SPM studies of water in the tip-sample gap where the roles of water sensing and instrument control are performed by the same physical interaction and where, in addition, that interaction is strongly or exclusively channeled via the water bridge. This means that while the presence/absence of a water bridge is readily detected there is generally limited scope for its detailed characterization on the nm scale. This critique applies to use of the normal force (sensed via measurement of amplitude, phase etc.) in atomic force microscopy [5], the shear force in dithered tip arrangements in near-field optical probe instruments [8, 9] and the tunnel current in STM as a means of detecting the water bridge. In the case of STM, for example, a current may persist over tens of nm of tip retraction [25] presumably until the connecting water ruptures. The water bridge is not only the medium for tunnelling at very small tip-sample separations but acts as a physical channel for all of the current of non-tunnelling origin [26] and there is no inherent instrumental distinction between these two sources.

## 2. Experiment

The experimental set-up comprises a STM (Digital Instruments Nanoscope E) housed in an environment controlled enclosure, that allows us to perform experiments under controlled RH and temperature conditions. The light output was collected by two 800 μm-core, low-OH⁻ content optical fibres, positioned ~1 mm from the junction at an angle of ~60° to the sample surface and with a mutual azimuthal angular separation of ~120°. The fibres were fed to a spectrometer (Acton Research SpectraPro 275 of focal length 275 mm) equipped with a charge coupled device camera (Andor DU420-OE) with a detection range of 400-1000 nm for recording the spectra. Both tips and samples were made from Au since this is the only metal of the requisite optical properties for good emission that may be used for extended periods under ambient conditions without contamination. Here we present results from two Au tips (T1 and T2) that were used to scan 30 nm thick polycrystalline Au samples. The tips were prepared by electrochemical etching of Au wire [27] and had end diameters of 24 nm (T1) and 170 nm (T2), estimated from scanning electron microscope (SEM) images. The Au thin films were thermally evaporated onto freshly cleaved mica substrates at ~$10^{-6}$ mbar base pressure and immediately transferred into the STM enclosure, maintained at the lowest relative humidity (RH ~ 7-10%). The films exhibit an average grain size of 30 nm and typically 1.2 nm rms roughness as measured by the STM. The enclosure RH level was allowed to stabilize for 30 minutes prior to the acquisition of data. All spectra were acquired in constant current mode of operation with $i_T$ = 10 nA at ±1.8 V sample bias. Each spectrum was obtained by averaging ten 60 s exposures. The spectral response of the detection system is incorporated in the modelling described presently.

## 3. Results and Discussion

Figures 2 and 3 show the evolution of the light emission spectra with RH for tips T1 and T2 at 10 nA tunnel and ±1.8 V sample bias. The two salient features are the decay of intensity (for both bias polarities) and the variation of the peak emission wavelength ($\lambda_p$) with RH. The decay of the peak intensity ($I_p$) with RH, plotted in Figure 4a-b, shows that the attenuation is faster for negative polarity than positive. Significantly, the emission is almost completely quenched for the highest RH only in the case of negative polarity for each tip. For positive bias the spectral peaks occur at $\lambda_p \approx$ 886 nm and 803 nm at the lowest humidity, for tips T1 and T2 respectively, and undergo a weak blue-shift with increasing RH, as indicated by the dashed lines in Figures 2a and 3a. A less pronounced blue-shift is observed for negative polarity. For the highest RH a marked red-shift is observed in the positive polarity but no peak position can be assigned in the negative polarity since the emission is quenched. Figure 3c shows the variation of $\lambda_p$ with RH.

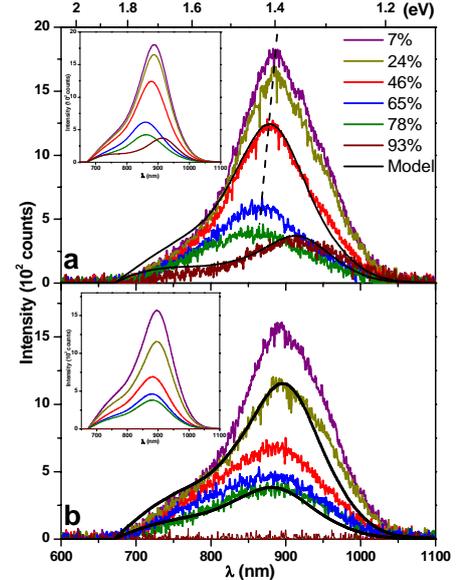

Figure 2: Light emission spectra for tip T1 with 10 nA tunnel current for (a) +1.8 V and (b) -1.8 V sample bias, at RH values indicated. The solid black lines are modeled spectra for selected values of RH; dashed line in (a) tracks shift in wavelength of peak intensity with increasing RH. Inset shows the modeled spectra over the entire range of RH.

In order to understand these observations it is necessary first to understand the physical basis of LESTM. This is achieved through the construction of an empirical model that satisfactorily replicates the experimental spectra. The model, retains essential elements of existing calculations for light emission from metal-oxide-metal [18, 19] and STM [12, 22-24] tunnel junctions, drawing on the work of Rendell and Scalapino (RS) [18] and Johansson [22] in particular. Various key physical factors in the emission process are described and summarized in Equations (1)–(3) and then brought together in Equation (4) which offers a description of the spectral output.

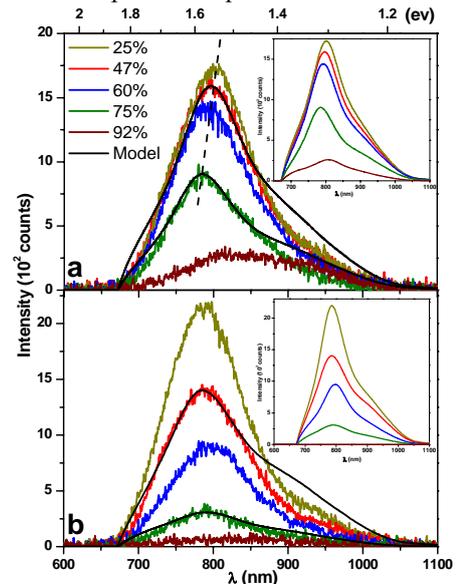

Figure 3: Light emission spectra for tip T2 with 10 nA tunnel current for (a) +1.8 V and (b) -1.8 V sample bias, at RH values indicated. The solid black lines are modeled spectra for selected values of RH; dashed line in (a) tracks shift in wavelength of peak intensity with increasing RH. Inset shows the modeled spectra over the entire range of RH.

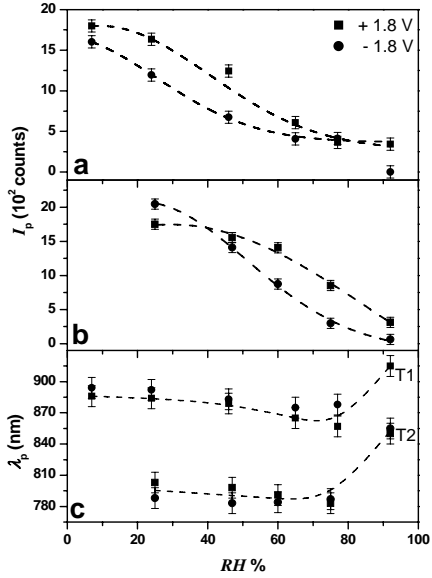

Figure 4: Decay of the primary peak emission intensity $I_p$ with increasing RH at ±1.8 V for (a) tip T1 and (b) tip T2. The dashed lines indicate a fitted Gaussian decay curve. (c) Variation of wavelength of peak intensity, $\lambda_p$, with RH for T1 and T2. The dashed lines are a guide to the eye.

In developing the schematic of LESTM process (Figure 1) in the tip sample nano-cavity, we follow the RS analysis in which the tip is envisaged as a sphere (radius $R$) held at a distance $d$ above a semi-infinite sample surface. It assumes that $d/R \ll 1$ (which is true in the context of STM) and that (i) the tip is a perfect conductor of dielectric constant $\varepsilon_2 \to -\infty$ (ii) the sample is a free electron material with $\varepsilon_1 = 1 - (\omega_P/\omega)^2$, $\omega_P$ being the bulk plasmon frequency and (iii) the surrounding medium has a real dielectric constant, $\varepsilon$. Under these assumptions the RS analysis yields the frequencies of the LSP modes for a particle-planar surface as:-

$$\omega_n = \omega_p \left[ \frac{\tanh(n+\tfrac{1}{2})\beta_0}{\varepsilon + \tanh(n+\tfrac{1}{2})\beta_0} \right] \quad n = 0, 1, 2, \ldots \quad (1)$$

where, $\beta_0 = \cosh^{-1}(1 + d/R)$. In order to calculate the radiated power we picture the charge oscillations due to optical frequency noise of the tunnel current as an Hertzian dipole source, of angular frequency $\omega$, located in the nano-cavity and oriented perpendicular to the surface. The power radiated at eigenfrequencies $\omega_n$ is then given by;

$$W \propto \frac{\omega^4}{(\omega_n^2 - \omega^2)^2 + \omega^2\gamma^2} \quad (2)$$

where, $\gamma$ denotes a damping term, empirically accounting for the spectral broadening of the emission peaks. The radiated power also depends on the source of the charge oscillations, the tunnel current, specifically its power spectral density [28],

$$C(\omega) \propto (1 - \hbar\omega/e|V_{bias}|); \quad \hbar\omega \leq V_{bias} \quad (3)$$

Finally, our previous LESTM experiments show that at fixed sample bias the emission intensity has an almost linear dependence on the tunnel current ($I_p \propto i_T^{1+\delta}$, $\delta \to 0$); also isochromat photon intensity is reported to have a $d$-dependence very similar to that of $i_T$ [12]. These observations lead us to incorporate a term $\propto \exp(-2\sqrt{2md}/\hbar)$ in the overall expression for the emission intensity:-

$$I = AK(\omega)\exp\left(\frac{2\sqrt{2md}}{\hbar}\right)\left(1 - \frac{\hbar\omega}{eV_{bias}}\right)\sum_{n=0}^{\infty} \frac{\omega^4}{(\omega_n^2 - \omega^2)^2 + \omega^2\gamma^2} \quad (4)$$

where, $K(\omega)$ is the experimental spectral response function and $A$ is an arbitrary scaling constant. A fit of Equation (4) to each experimental spectrum yields the parameters $d$, $R$, $\gamma$, and $\varepsilon$. However, before proceeding to the use of Equation (4) in modelling the spectra, it is necessary to consider the critical nanoscale parameters that characterize the tip-sample cavity both geometrically and optically. Firstly, the value of tip radius, $R$, in the modelling is constrained by independent measurement from SEM micrographs ($R_{SEM}$). Secondly, from previous spectral modelling at low RH conditions (the starting point in the process) the value of $d$ is found to be in the range 0.50 to 0.65 nm; this scale of tunnel gap is highly consistent with that used in analysis of LESTM results in the literature [23]. Optically, the LSP mode volume forms the third critical size parameter. The LSPs are confined to a volume specified by a lateral confinement length, $L_n$, given approximately by [18]

$$L_n = \sqrt{Rd/(2n+1)}; \quad (n+\tfrac{1}{2})\beta_0 \ll 1 \quad (5)$$

$L_n$, decreases with increasing LSP energy and defines the volume of resonant tip-sample electromagnetic coupling or the 'optically active' nano-cavity.

The nano-cavity undergoes significant changes with increasing RH since a sharp tip in close proximity to the sample surface ($d < 1.0$ nm) acts as a nucleation site causing capillary condensation in the tip-sample junction forming a water bridge between them; the radius of this bridge is the fourth critical size parameter. Capillary condensation is a well-known effect that relates to the condensation of liquid vapor between two solid surfaces in close proximity [29]. Under equilibrium conditions, if we assume the water bridge approaches a cylindrical geometry as depicted in Figure 1b, then the radius of the water bridge is given by the Kelvin radius, $r_K$:-

$$r_K \propto -\frac{\sigma V_M}{R_G T \ln(RH)} \quad (6)$$

where $\sigma$ is the surface tension, $V_M$ is the molar volume, $R_G$ is the universal gas constant and $T$ is the temperature. The factor, $\sigma V_M / R_G T$ (= 0.525 nm for water at 300 K), essentially sets the scale for Equation (6) and, at fixed temperature, $r_K$ is a function of RH only. Although this equation derives from continuous medium theory and uses the macroscopic surface tension, $\sigma$, it has been verified experimentally down to $r_K \sim 3$ nm [30]. Moreover, theoretical simulation, estimates the minimum meniscus radius for condensed water in a tip-sample junction as 0.85 nm [31], setting a lower limit to the applicability of the Kelvin equation.

A build-up of water in the nano-cavity will change the LSP modal frequencies through an increase in the effective optical-frequency dielectric constant which we will define here as:-

$$\varepsilon_{eff} = (1.0 + 0.77v_{water}) \quad (7)$$

where $v_{water} \sim (r_K/L_n)^2$ is the volume fraction for a cylindrical water bridge assuming the bulk dielectric constant of 1.77 for water. If this simple picture of water in the cavity is valid there should then be consistency between the values of $\varepsilon_{eff}$ derived from Equation (7) and those for $\varepsilon$ resulting from the spectral modelling using Equation (4). The link connecting these two quantities is the confinement length $L_n$ (Equation (5)) which arises from the same LSP modal analysis on which Equation (1) is based.

The condensation of water in the nano-cavity has major consequences electronically as well as optically. The central thesis is that the decay in light emission intensity is due to an increase in the tunnel gap, $d$, (Equation (4)) as $r_K$ grows with increasing RH. Previous investigations with STM at elevated RH confirm that $d$ is notably larger [25, 26, 32-34] than under normal operation, as can be the current [35]. Clearly tunnelling is not the mechanism of current flow at large $d$ (> 4 nm) and is more likely of electrochemical origin. The total tip–sample current can be thus written as $i_{total} = i_T + i_{alt}$, $i_{alt}$ representing current flow via *various* possible alternative charge transfer pathways other than by direct metal-metal tunnelling, $i_T$. Presence of a polar and ionizable electrolyte, i.e. water, in the junction would facilitate these alternate routes of charge transfer, more so under the high electric field (~ $10^9$ V/m) in the cavity that quite likely ionizes water to a certain extent. The alternative charge transfer routes manifest as (i) 'intermediate states' [32] lowering the effective barrier height or (ii) electrochemical processes [26]. The crucial point in the present context is that any charge transfer mechanism, other than tunnelling would not lead to excitation of LSPs and thus light emission. An increase in the water bridge radius (with RH) and consequently the electrode/water interface area would lead to an increase in $i_{alt}$. Given the constant-current operation mode of the STM, $i_{total}$ increases causing the feedback to increase $d$ to reduce $i_T$ in order to keep $i_{total}$ constant. This is the overriding effect since $i_{alt}$ is only weakly dependent on $d$ [26], in contrast to the exponential decay of $i_T$ with $d$. This effect is depicted schematically in Figure 1b.

An increase in $d$ also leads to a blue shift of the emission through Equation (1), but a more pronounced blue shift than that observed experimentally. This is countered by a red–shift brought about through an increase in the average dielectric constant of the gap, $\varepsilon$, with increasing RH (Equation (1)). In summary, an increase in RH and thus in the water fill factor relative to the LSP modal volume brings about these two opposing effects which, in combination, give rise to the evolution of the spectra with RH as observed in Figures 2 and 3.

Modelling of the spectra starts with the lowest humidity case and proceeds as follows. The best fit to the lowest RH (7%) experimental data for tip T1 (for positive polarity) yields $R \sim 9$ nm ($R_{SEM} = 12$ nm), $\gamma = 31$ rad$^{-2}$ s$^2$, $d = 0.60$ nm and $A = 480$, with $\varepsilon = 1.00$. The calculated spectrum (inset of Figure 2a, purple curve) shows emission corresponding to LSP mode $n = 0$ with a peak at 880 nm. Similarly, the best fit values (for the lowest-RH spectrum) for tip T2, for positive bias yield $R = 87$ nm ($R_{SEM} = 85$ nm), $\gamma = 28$ rad$^{-2}$ s$^2$, $d = 0.9$ nm and $A = 370$ with $\varepsilon = 1.03$. The calculated spectrum (inset of Figure 3a, olive green curve) shows emission corresponding to LSP mode $n = 1$ with a peak at 790 nm for RH = 25%. The values of $R$, $\omega_p$, (5.47 x $10^{15}$ rad s$^{-1}$) and $\gamma$ were kept fixed at the best-fit values for the lowest RH while only $d$ and $\varepsilon$ were changed [36] to fit subsequent spectra at higher RH. The exercise is repeated across the four sets of spectra and the calculated spectra are shown in the insets of Figures 2 and 3 for both polarities. Fits to selected spectra are also plotted with the experimental data for direct comparison. Figure 5 shows a plot of the fit parameters $d$ and $\varepsilon$ with RH for tip T1. The positive polarity results show that the spectral evolution with increasing RH requires an increase in both $d$ and $\varepsilon$, as explained above. The interesting feature about $\varepsilon$ is that there is very good agreement with $\varepsilon_{eff}$ (Figure 5), evaluated on the basis of the Kelvin radius model, combined with the simple calculation based on the partial filling of the active nano-cavity (Equations (5)-(7)). Note the extreme sensitivity of light emission to water in the gap – the large changes of Figures 2 and 3 take place within a vertical range ($\Delta d$) of ~1.5 nm and within a lateral range set by $2L_n$ where the confinement lengths for the dominant emitting modes are $L_0$ ~ 2.3 nm and $L_1$ = 3.0 nm for T1 and T2 respectively; simultaneously we also obtain an average optical characterization on the gap dielectric environment. The general trend of increasing $\varepsilon$ and $d$ for the negative polarity and its physical understanding qualitatively follows that for the positive polarity. However, quantitatively, there are differences in $\varepsilon$ and $d$ for the two polarities (Figure 5) that essentially characterize a fundamental asymmetry in the system.

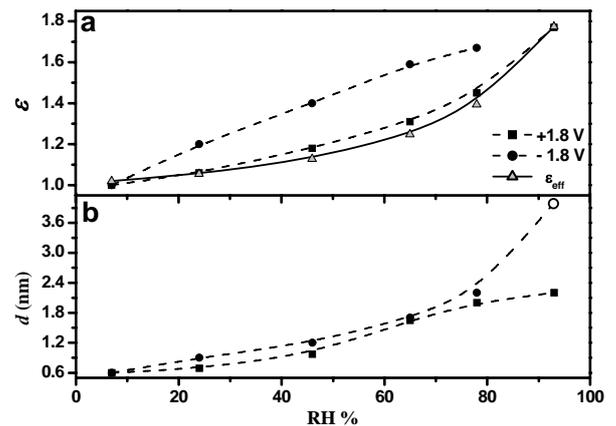

Figure 5: Parameters (a) $\varepsilon$ and (b) $d$, generated by Equation (4) for fitting spectra in Figure 1a-b. The open circle in (b) at the highest RH (for -1.8 V bias) indicates the minimum value of $d$ necessary to reduce the calculated emission to within detector noise level. The grey triangles in (a) indicate the values of $\varepsilon_{eff}$ calculated from Equation (7).

Previous X-ray studies [37] indicate that the origin of this asymmetry lies with the structural arrangement of polar water molecules under the influence of a strong **E**-field in an asymmetric geometry. A lower areal density or more open structure of polarized water molecules was found on a negatively charged sample where the hydrogen atoms are the species immediately adjacent to the surface, in contrast to water molecules on a positively charged surface. Originating with the different arrangement of molecules for the two polarities, we suggest that conformity to the simplified picture of a cylindrical geometry for the water bridge differs for the two cases. There is very good conformity for positive polarity and molecular dynamics simulations [10] indicate the formation of a cylindrical column under similar high **E**-field conditions. However, for negative polarity if the arrangement of water molecules is associated with a concave meniscus profile (especially near the sample surface) at a given RH, and a consequently larger value for $i_{alt}$ (compared to positive polarity) then the modelling will generate a greater value of tip-sample separation, $d$, than is actually the case. The effects of this have to be retrieved by a corresponding increase in $\varepsilon$ in the spectral modelling, higher than those based on equation 7. Finally, the most marked polarity-dependent contrast comes at the highest RH. For positive polarity the effect of increasing $\varepsilon$ ($\varepsilon_{eff}$) fully to 1.77 dominates the effect of further increase in $d$ and yields a sudden, pronounced red shift. In the negative polarity complete quenching of emission occurs resulting in a large value of $d$ ($\geq 4$ nm), for which the tunnelling probability is infinitesimally small. Experimentally, we observe an unstable feedback going into spontaneous oscillations (unobserved for positive bias), probably associated with the intermittent formation and breakup of the water bridge, resulting in complete quenching of emission.

The understanding of the asymmetry in the results in terms of the behaviour of polar water molecules in a high, asymmetric **E**-field is corroborated by the LESTM results of Figure 6 where an inert, non-polar dielectric fills the nano-cavity (a 5-ring polyphenyl ether (PPE) [38] which is optically transparent to visible and near infra-red light with $\varepsilon \sim 2.657$ was used). Importantly, the inset of Figure 6 shows the light emission with PPE to be virtually independent of polarity in contrast to that in presence of water. Figure 6 demonstrates the consequences of a change in nano-cavity $\varepsilon$ alone. The spectrum from clean Au (upper black curve) shows four peaks at 975, 880, 790 and 720 nm. The corresponding calculated spectrum (red) shows matching peaks with mode numbers $n = 2, 3, 4$ and 5 and best-fit parameters $d = 0.58$ nm, $R = 405$ nm, $\varepsilon = 1.0$, $\gamma = 40$ rad$^{-2}$s$^2$ and $A = 46$.

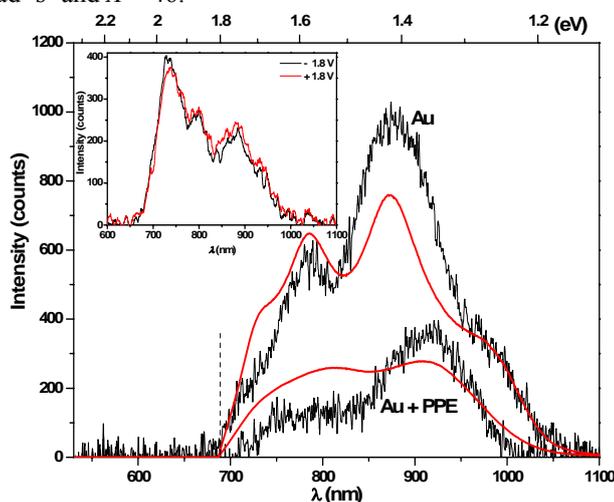

Figure 6: Experimental emission spectra from bare Au tip/Au sample at RH = 10% in air and from PPE covered Au surface taken with $V_{bias}$= +1.8 V and 10 nA tunnel current. The modeled spectra are shown red in the main graph. Inset shows the spectra from PPE covered Au surface taken with $V_{bias}$ = +1.8 V and -1.8 V.

The experimental spectrum for the case where PPE fills the gap (lower black curve) exhibits a distinct peak at 925 nm and a broad shoulder at lower wavelengths while the corresponding, calculated spectrum is generated by changing $\varepsilon$ to 2.657 (best fit case was actually $\varepsilon = 2.60$), keeping all other parameters constant. It shows a peak at 939 nm with two smaller peaks at 804 nm and 719 nm replicating the spectral shoulder of the experiment. (While the simple model developed here gets the core spectral information essentially correct there is a shortcoming in describing relative peak intensities from different LSP modes.)

## 4. Conclusion

In conclusion, it has been shown that LESTM acts as a highly sensitive means of characterizing the dielectric properties of the medium in the tip-sample gap, on a volume scale ranging from sub-zepto- to atto-liter ($<10^{-21}$ to$10^{-18}$ l); light emission can therefore offer a useful additional channel of information from STM in aqueous and other fluid environments. Specifically, the intensity decay and spectral shift in the emitted light with increasing relative humidity offers a means of tracking the capillary condensation of water in the gap on the (sub-) nm scale. The results are explained through the development of an empirical model that successfully describes the changes in emission as due to progressive water filling of the gap or, more pertinently, in optical terms, of the LSP modal volume. Asymmetry in the results with respect to polarity of the applied bias is due to the polar nature of the water molecules; this observation is corroborated by the absence of such asymmetry when a medium of non-polar molecules occupies the gap. While water bridging of the gap is of fundamental importance in all forms of ambient SPM, the results presented here are of particular relevance to apertureless near-field scanning probe microscopy and tip enhanced Raman scattering. The distinctive nature of this approach, relative to other SPM studies of the capillary condensation of water in the tip-sample gap, lies in the fact that the physical entity (the emitted light) used to probe and characterize the gap medium is not used as the feedback parameter for instrument control. The modal volume of the originating plasmon excitation sets an upper limit for the volume of the water bridge that may be probed, but below this limit the growth of the water bridge may be characterized with very high sensitivity.

**Acknowledgements.** The authors acknowledge financial support from the UK Engineering and Physical Sciences Research Council (EPSRC-EP/D048850/1) and 'Nanotec Northern Ireland' supported by EC funding through Invest Northern Ireland.